# Simple methods to mesure an intensity modulator efficiency versus the frequency of its driving signal

Gilles Feugnet
Thales Research and Technology
Palaiseau, France
gilles.feugnet@thalesgroup.com

Salma Essawab
Thales Research and Technology
Palaiseau, France
salma.essawab@thalesgroup.com

Jimmy Pennanech
Thales Research and Technology
Palaiseau, France
jimmy.pennanch@thalesgroup.com

Delphine Pommier
Thales Research and Technology
Palaiseau, France
delphine.pommier@thalesgroup.com

Jérome Bourderionnet
Thales Research and Technology
Palaiseau, France
jerome.bourderionnet@thalesgroup.com

Arnaud Brignon
Thales Research and Technology
Palaiseau, France
arnaud.brignon@thalesgroup.com

Benoit Benazet
Thales Alenia Space
Toulouse, France
benoit.benazet@thalesgroup.com

Homa Zarebidaki
Centre Suisse d'Electronique
et de Microtechnique
Neufchatel, Swiss
homa.zarebidaki@csem.ch email

Waldemar Schmidt
Rosenberger Hochfrequenztechnik
GmbH & Co. KG,
Fridolfing, Germany
waldemar.schmidt@rosenberger.com

***Abstract*—We describe and compare in this paper two common methods to measure the half-wave voltage (Vπ) of an electro-optical modulator versus RF frequency of the driving electronic signal. We detail the required calibration procedures and demonstrate the methods experimentally from 100 MHz up to 40 GHz. Guidelines are finally given to select the most appropriate method depending on the application and modulator characteristics.**

## I. Introduction

Intensity Mach-Zehnder Modulators (MZM) are widely used in photonics and optical communication systems. Their efficiencies are characterized by the half-wave voltage Vπ or expressed in a more complex Figure of Merit[1] depending on it. Vπ is the voltage required to induce a phase shift of π radians between the arms of the modulator and it depends on the frequency of the modulation signal. This paper proposes two methods to measure how Vπ changes versus these frequencies. Section II is dedicated to derive general equations supporting our analysis, including the extinction ratio of the modulators. Section III and IV detail methods based on the detection of the modulated light respectively with a power meter or with an Optical Spectrum Analyser (OSA) and emphasize the calibration procedure to ensure the relevance of the calculated Vπ. Validation based on experiments results is exposed in Section Part V. We conclude by comparing our methods to existing ones and giving some clues on when to use them and extend the discussion to phase modulators.

## II. Expression of the amplitude and intesity of the ligth after an intensity modulator

In our discussion, we are considering modulators driven in push-pull mode, i.e. with opposite phases $\varphi_1 = -\varphi_2 = \varphi$ in the arms 1 and 2 of the modulator but our analysis can be adapted to other cases based on [2]. To include the case of modulators with unbalanced in and out couplers, we will write the intensity coupling ratio $\kappa$ as $\kappa = 1/2 - \alpha$ with $\alpha \neq 0$. With our convention, the electric field after a modulator with two 1×2 MMI [1] in/out couplers can be written as:

$$E_{out} = E_{in}\, e^{-j\omega_0 t}\left[(1-\kappa)e^{-j\varphi} + \kappa e^{j\varphi}\right] \quad (1)$$

$$= E_{in}\, e^{-j\omega_0 t}\left[\left(\frac{1}{2}+\alpha\right)e^{-j\varphi} + \left(\frac{1}{2}-\alpha\right)e^{j\varphi}\right]$$

$$= E_{in}\, e^{-j\omega_0 t}\left[cos(\varphi) - 2j\,\alpha sin(\varphi)\right] \quad (2)$$

Introducing $\omega_{RF}$ and $V_{RF}$, respectively the RF frequency and RF peak voltage of the modulation signal and $\varphi_{DC}$ a static phase difference (due for instance to an unbalanced of lengths of the MZM arms or a DC bias added to the modulation) and the peak voltage $V_{\pi_{RF}}$, we can write that

$$\varphi = \frac{\pi}{2}\frac{V_{RF}}{V_{\pi_{RF}}} sin(\omega_{RF}t) + \varphi_{DC} \quad (3)$$

that swings over $\pi$ radians for $V_{RF} = V_{\pi_{RF}}$ since then $-\frac{\pi}{2} \leq \frac{\pi}{2}\frac{V_{RF}}{V_{\pi_{RF}}} sin(\omega_{RF}t) = \frac{\pi}{2} sin(\omega_{RF}t) \leq \frac{\pi}{2}$.

Using the trigonometric formulas to develop cosine and sine of $\frac{\pi}{2}\frac{V_{RF}}{V_{\pi_{RF}}} sin(\omega_{RF}t) + \varphi_{DC}$ and the Jacobi-Anger expansion[4]

$$cos\big(a\,sin(\omega_m t)\big) = J_0(a) + 2\sum_{k=1}^{+\infty} J_{2k}(a) cos(2k\omega_m t)$$

and

$$sin\big(a\,sin(\omega_m t)\big) = 2\sum_{k=1}^{+\infty} J_{2k-1}(a) sin\big((2k-1)\omega_m t\big)$$

shows that the electric field has three types of components:

- a DC component

$$E_{in}e^{j\omega_0 t} J_0\left(\frac{\pi}{2}\frac{V_{RF}}{V_{\pi_{RF}}}\right)\big(cos(\varphi_{DC}) - 2j\alpha sin(\varphi_{DC})\big) \quad (4a)$$

- a component with only the even harmonics of the modulation frequency

$$2E_{in}e^{j\omega_0 t}\big(cos(\varphi_{DC}) - 2j\alpha sin(\varphi_{DC})\big)\sum_{k=1}^{+\infty} J_{2k}\left(\frac{\pi}{2}\frac{V_{RF}}{V_{\pi_{RF}}}\right) cos(2k\omega_m t)$$

$$= E_{in}e^{j\omega_0 t}\left(cos(\varphi_{DC}) - 2j\alpha sin(\varphi_{DC})\right) \sum_{k=1}^{+\infty} J_{2k}\left(\frac{\pi}{2}\frac{V_{RF}}{V_{\pi_{RF}}}\right)\left(e^{j2k\omega_m t} + e^{-j2k\omega_m t}\right) \quad (4b)$$

- a component with only the odd harmonics of the modulation frequency

$$-2E_{in}e^{j\omega_0 t}\left(sin(\varphi_{DC}) + 2j\alpha\, cos(\varphi_{DC})\right)\sum_{k=1}^{+\infty} J_{2k-1}\left(\frac{\pi}{2}\frac{V_{RF}}{V_{\pi_{RF}}}\right) sin\left((2k-1)\omega_m t\right)$$
$$= jE_{in}e^{j\omega_0 t}\left(sin(\varphi_{DC}) + 2j\alpha\, cos(\varphi_{DC})\right) \sum_{k=1}^{+\infty} J_{2k-1}\left(\frac{\pi}{2}\frac{V_{RF}}{V_{\pi_{RF}}}\right)\left(e^{j(2k-1)\omega_m t} - e^{-j(2k-1)\omega_m t}\right) \quad (4c)$$

with the decomposition in complex exponentials giving the amplitude and frequencies of the upper and lower sidebands. From equation (1), the intensity is given by

$$I_{out}/I_0 = \left|(1-\kappa)e^{-i\varphi} + \kappa e^{i\varphi}\right|^2$$
$$= 1 - 2\kappa(1-\kappa) + 2\kappa(1-\kappa)\, cos(2\varphi)$$
$$= \frac{1+\varepsilon}{2} + \frac{1-\varepsilon}{2}\, cos\left(\pi\frac{V_{RF}}{V_{\pi_{RF}}} sin(\omega_m t) + 2\varphi_{DC}\right) \quad (5)$$

by introducing the extinction ratio

$$\varepsilon = \frac{I_{min}}{I_{max}} = 1 - 4(1-\kappa)\kappa.$$

Using the same Jacobi-Anger expansion, it is possible to show that the intensity has three types of components:

- a DC component

$$I_{DC} = \frac{I_0}{2}\left[(1+\varepsilon) + (1-\varepsilon)J_0\left(\pi\frac{V_{RF}}{V_{\pi_{RF}}}\right) cos(2\varphi_{DC})\right] \quad (6a)$$

- a component with only the even harmonics of the modulation frequency

$$I_{odd} = I_0 cos(2\varphi_{DC})(1-\varepsilon)\sum_{k=1}^{+\infty} J_{2k}\left(\pi\frac{V_{RF}}{V_{\pi_{RF}}}\right) cos(2k\omega_m t) \quad (6b)$$

- a component with only the odd harmonics of the modulation frequency

$$I_{odd} = -I_0 sin(2\varphi_{DC})(1-\varepsilon)\sum_{k=1}^{+\infty} J_{2k-1}\left(\pi\frac{V_{RF}}{V_{\pi_{RF}}}\right) sin\left((2k-1)\omega_m t\right) \quad (6c)$$

## III. Measurement with an optical power meter

A power meter measures only the DC components (6a) of the intensity and thus its reading P is given by

$$P = \frac{P_0}{2}\left[(1+\varepsilon) + (1-\varepsilon)J_0\left(\pi\frac{V_{RF}}{V_{\pi_{RF}}}\right) cos(2\varphi_{DC})\right] \quad (7)$$

where $P_0$ is the power meter reading without RF ($J_0\left(\pi\frac{V_{RF}}{V_{\pi_{RF}}}\right) = 1$) and with the modulator biased to maximize its output power ($cos(2\varphi_{DC}) = 1$).

The MZM is biased at extinction for the measurement step ($cos(2\varphi_{DC}) = -1$) without RF so that

$$P = \frac{P_0}{2}\left[(1+\varepsilon) - (1-\varepsilon)J_0\left(\pi\frac{V_{RF}}{V_{\pi_{RF}}}\right)\right]$$

In that case, for low $V_{RF}/V_{\pi_{RF}}$, $J_0\left(\pi\frac{V_{RF}}{V_{\pi_{RF}}}\right) \approx 1 - \left(\frac{\pi}{2}\frac{V_{RF}}{V_{\pi_{RF}}}\right)^2$ from [5] so

$$P \approx \frac{P_0}{2}\left[2\varepsilon + (1-\varepsilon)\left(\frac{\pi}{2}\frac{V_{RF}}{V_{\pi_{RF}}}\right)^2\right] \approx \frac{P_0}{2}\left(\frac{\pi}{2}\frac{V_{RF}}{V_{\pi_{RF}}}\right)^2$$

if the extinction ratio $\varepsilon$ is close to zero. This shows that very weak variations can be measured but also explains why it is very important to measure $P_0$. If $P_0$ is not precisely measured, the fitting parameter is not $V_{\pi_{RF}}$ but $V_{\pi_{RF}}/\sqrt{P_0}$ and meaningless value of $V_{\pi_{RF}}$ can be calculated depending on the choice of $P_0$. Once $P_0$ is measured, the measurement is performed at a given modulation frequency $\omega_m$ by varying the applied peak voltage $V_{RF}$ and fitting the curve $P(V_{RF})$ with only $V_{\pi_{RF}}$ as an adjustable fitting parameter.

An optional faster method, for modulators with low enough $V_{\pi_{RF}}$, is to reach the maximum output power, $\frac{P_0}{2}(1+\varepsilon)$, corresponding of the first zero of $J_0$ with $V_{\pi_{RF}} = \pi\frac{V_{RF}}{2.4}$ (8).

## IV. Measurement with an optical spectrum analyser

An optical spectral analyzer (OSA) allows measuring the optical power in the sidebands created by the MZM modulation. The set of equations 4 leads to following respective expression of the powers of the optical carrier, the first and second sidebands:

$$P_{OSA,carrier} = A\, J_0^2\left(\frac{\pi}{2}\frac{V_{RF}}{V_{\pi_{RF}}}\right)\left(cos^2(\varphi_{DC}) + 4\alpha^2 sin^2(\varphi_{DC})\right)$$
$$P_{OSA,\omega_m} = A\, J_1^2\left(\frac{\pi}{2}\frac{V_{RF}}{V_{\pi_{RF}}}\right)\left(sin^2(\varphi_{DC}) + 4\alpha^2 cos^2(\varphi_{DC})\right)$$
$$P_{OSA,2\omega_m} = A\, J_2^2\left(\frac{\pi}{2}\frac{V_{RF}}{V_{\pi_{RF}}}\right)\left(cos^2(\varphi_{DC}) + 4\alpha^2 sin^2(\varphi_{DC})\right)$$
$$P_{OSA,3\omega_m} = A\, J_3^2\left(\frac{\pi}{2}\frac{V_{RF}}{V_{\pi_{RF}}}\right)\left(cos^2(\varphi_{DC}) + 4\alpha^2 sin^2(\varphi_{DC})\right)$$

The coefficient A depends on the output power of the modulator, the coupling efficiency of the light in the OSA, its sensitivity, its filter shape/bandwidth… As with the power meter method, measuring A is important as for small $\frac{V_{RF}}{V_{\pi_{RF}}}$ [5] we have $J_1^2\left(\frac{\pi}{2}\frac{V_{RF}}{V_{\pi_{RF}}}\right) \approx \left(\frac{\pi}{2}\frac{V_{RF}}{V_{\pi_{RF}}}\right)^2$ so

$$P_{OSA,\omega_m} \approx A\left(\frac{\pi}{2}\frac{V_{RF}}{V_{\pi_{RF}}}\right)^2\left(sin^2(\varphi_{DC}) + 4\alpha^2 cos^2(\varphi_{DC})\right)$$

which means that the fitting parameter would be $V_{\pi_{RF}}/\sqrt{A}$ if A is unknown. A corresponds indeed to the carrier power measured by the OSA, $P_{OSA,carrier}\,(no\ RF, max)$, with no RF and with the modulator biased for maximum output power since then $J_0^2\left(\frac{\pi}{2}\frac{V_{RF}}{V_{\pi_{RF}}}\right) = 1$, $cos^2(\varphi_{DC}) = 1$ and $sin^2(\varphi_{DC}) = 0$. Once this value is measured, measurements of the power of the first sidebands can be performed by first setting $sin^2(\varphi_{DC}) = 1$ i.e. by biasing the modulator at extinction without RF modulation and then varying $V_{RF}$ and fitting the curves:

$$P_{OSA,\omega_m} = P_{OSA,carrier}(no\ RF, max)\, J_1^2\left(\frac{\pi}{2}\frac{V_{RF}}{V_{\pi_{RF}}}\right)$$

When biasing the modulator for maximum output power to measure the coefficient A ($cos^2(\varphi_{DC}) = 1$) and when biasing it at extinction to measure the powers of the first sidebands versus $V_{RF}$, ($sin^2(\varphi_{DC}) = 1$), contributions from the finite extinction ratio disappear which means that this method is well suited even for modulators with low extinction as this might happen during the development of a new modulator technologies.

A complimentary analysis, to avoid the calibration step, can be conducted by cross-comparing the ratio of power between the optical carrier components and the optical sidebands or between the sideband themselves. The biasing will depend on which sidebands are considered.

Again, an optional faster measurement would be to drive the modulator to the maximum power of the first sidebands, $P_{OSA,\omega_m}^{max}$, reached for the first maximum of $J_1$ i.e. $\frac{\pi}{2}\frac{V_{RF}}{V_{\pi RF}} = 1.83$ or fit

$$P_{OSA}(k=1) = P_{OSA,\omega_m}^{max}\frac{J_1^2\left(\frac{\pi}{2}\frac{V_{RF}}{V_{\pi_{RF}}}\right)}{J_1^2(1.83)} = \frac{P_{OSA,\omega_m}^{max}}{0.34}J_1^2\left(\frac{\pi}{2}\frac{V_{RF}}{V_{\pi_{RF}}}\right)$$

for a more precise result than only a single value. One can note that this condition leads to lower RF power compare to the condition (8).

## V. Experimental results

To validate and compare our methods, we use a RIO laser (10 mW) and an Exail 40GHz intensity modulator. The output of the modulator is split into two beams: one beam connected to a Thorlabs power meter with a S122C head and the other beam connected to an APEX AP2083A OSA set to 20MHz resolution. All the measurements with the power meter and the OSA are done simultaneously. Since they rely on sweeping the applied peak voltages, it is first necessary to calibrate the RF source, in our case a Rohde & Schwarz ZVA VNA set to deliver only one RF frequency at a time (but a cheaper RF generator would be suitable). To measure the real RF power delivered to the modulator, we connect the output of the VNA to a Rohde & Schwarz RF power meter NRP67T. The same RF cable in the same positioning between the VNA and the power meter, is used when characterizing the modulator. This is very important as RF cables are lossy and changing the cable might affect delivered powers. Correction for the potential losses of the RF cable is of course possible but de-embedding remains a tricky process ([6],[7],[8]) still subject to research[9]. Besides, the modulator is connected to a cable anyway when integrated in a system, so once the cable is chosen for whatever reason (flexibility, bandwidth, cost…), it is better to perform the measurements with it as the Device Under Test is indeed the modulator and its cable. In Fig.1, we plot the real RF power versus the RF frequency after a 1hour warm-up time of the RF power meter.

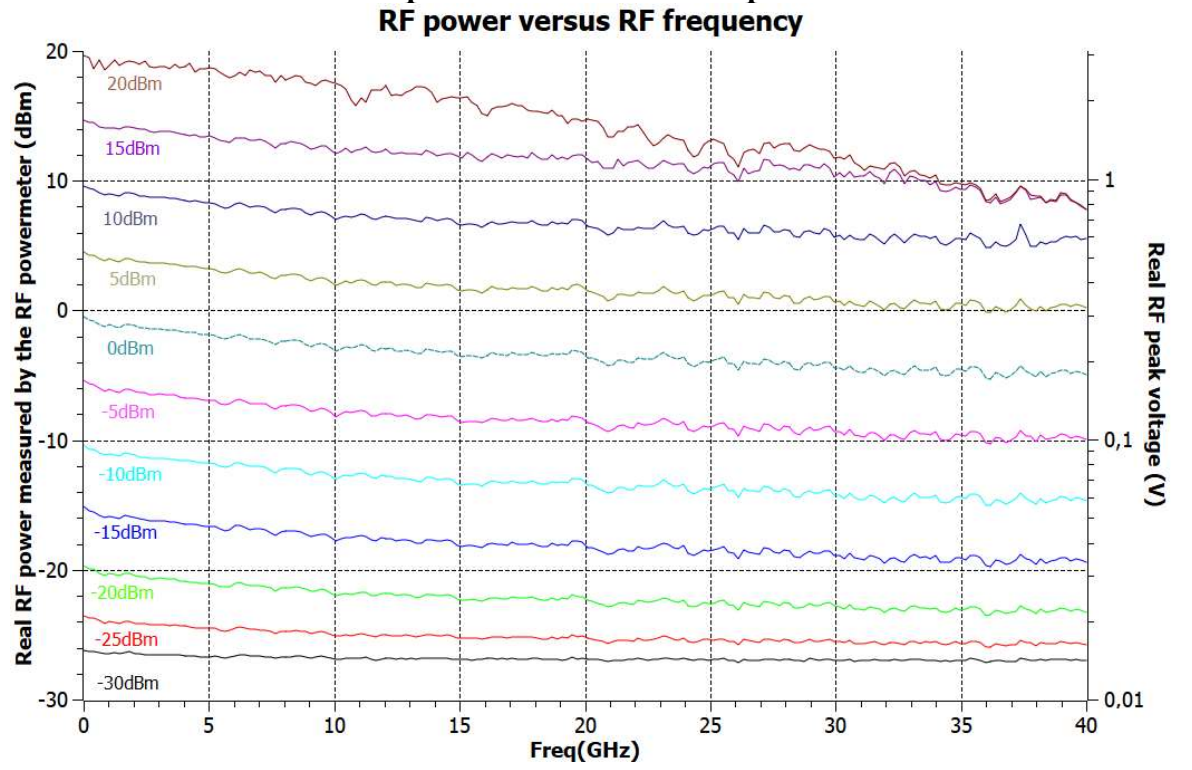

Fig.1: Calibration of the RF source. Displayed RF powers on the VNA are indicated at the bottom of the corresponding curves. The left vertical scale is the real power measured with the RF power meter. The right vertical scale is the corresponding peak voltage scale, calculated using equation 9.

As can be seen, the important discrepancies between expected/set and real powers justify our calibration step. We did not use the VNA power-calibrated mode as considering Fig.1, it might limit the frequency range at high RF power.

We believe the calibrations of the optical parameters $P_0$,$\varepsilon$, $P_{OSA,carrier}(no\ RF, max)$ are rather straightforward (compared for example to the calibration of a fast photodetector when an Electrical Spectrum Analyzer is used as described in [10]). In order to make sure these parameters do not drifts during the measurements (the APEX OSA has a very good resolution but is rather slow), we checked them regularly when changing the RF frequency. For the sake of clarity, we summarize in the Table 1 the procedures to follow that are similar for both methods.

| | Measure with a power meter | Measure with an OSA |
|---|---|---|
| Calibration | Set MZM for maximum transmission. No RF. | |
| | Measure optical power $P_0$ and extinction ratio $\varepsilon = \frac{I_{min}}{I_{max}}$ | Measure optical power of the carrier $P_{OSA,carrier}(no\ RF, max)$ |
| Measurements | Set MZM to extinction. No RF. Once done, apply RF power | |
| | Record optical power at the output of the modulator | Measure optical power of the first sidebands with OSA |
| | Optional: $V_{RF}$ for maximum output power $V_{\pi_{RF}} = \pi\frac{V_{RF}}{2.4}$ | Optional: $V_{RF}$ for maximum power of the first sidebands $V_{\pi_{RF}} = \frac{\pi}{2}\frac{V_{RF}}{1.83}$ |

Table 1: summary of the calibration and measurement steps.

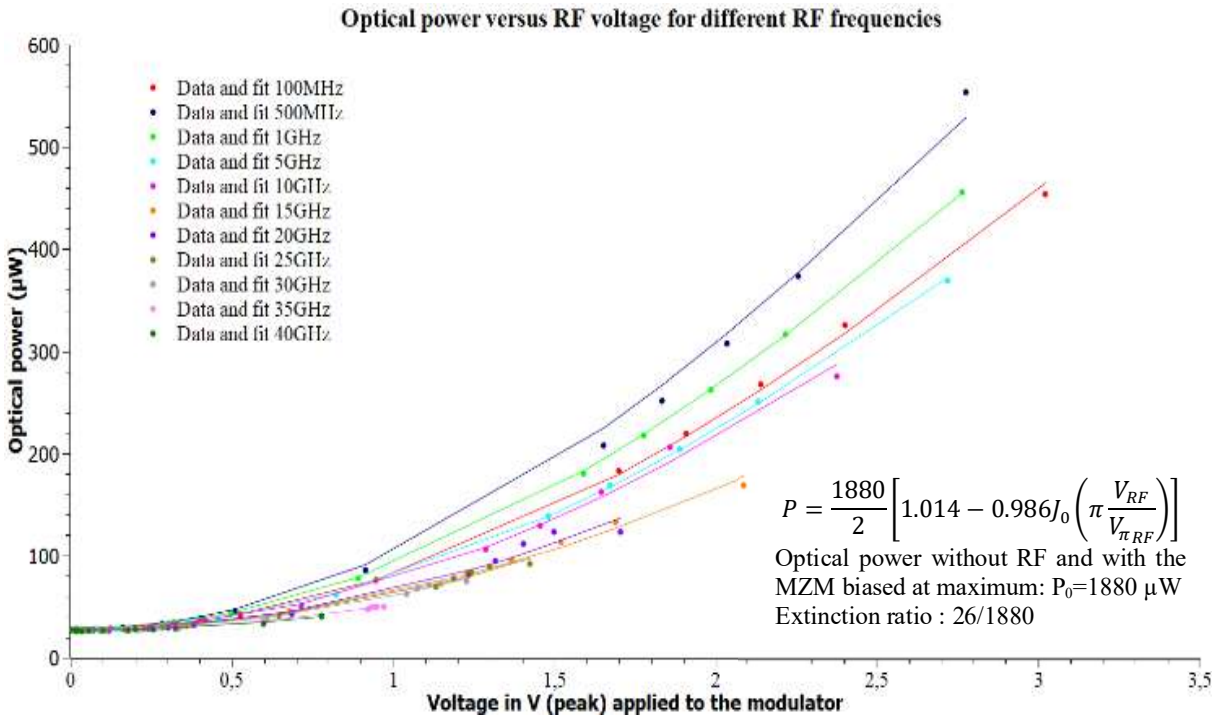

Fig.2. Experimental data(dots) and fits (lines) of the optical powers measured by the power meter for modulation frequencies from 100MHz to 40GHz. $P_0$, $\varepsilon$ and the fitting function are indicated in the inner right bottom insert.

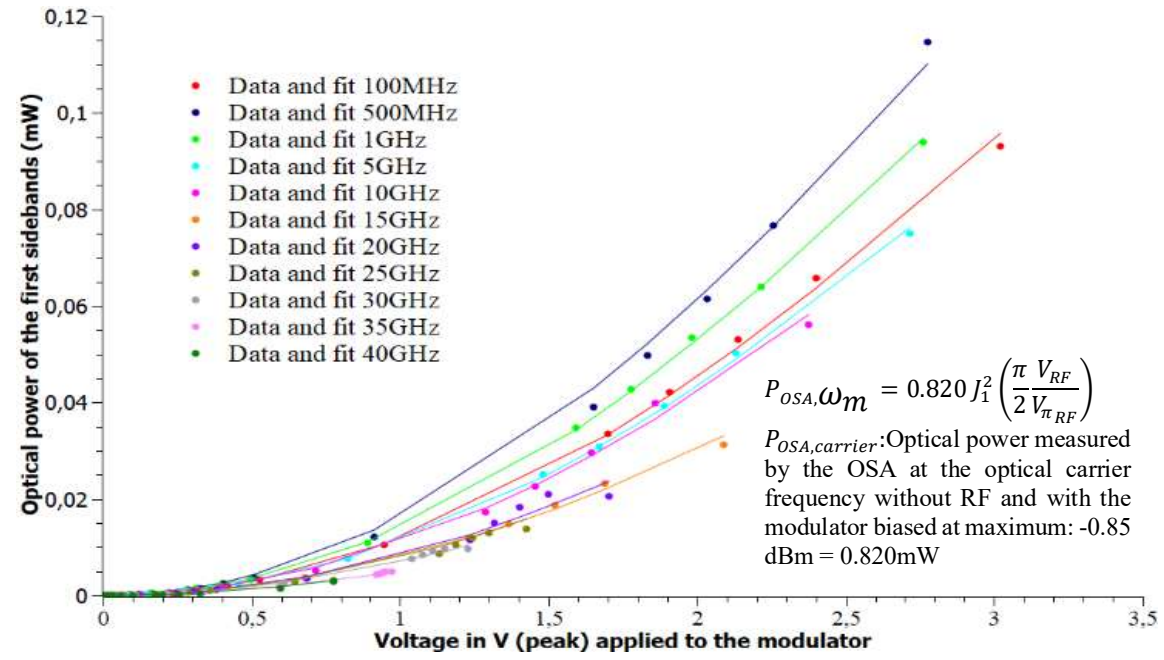

Fig.3. Experimental data(dots) and fits (lines) of the optical powers of the first sideband measured with an OSA for modulation frequencies from 100MHz to 40GHz.

Fig.2 shows the measured optical powers and the fitted curves versus applied peak voltage $V_{RF}$ using the measured values of $P_0$ and $\varepsilon$ as defined in section III (there are indicated in the inset). In Fig.3, we display the same information with the

OSA measurements. To calculate the applied peak voltage $V_{RF}$, we use the relationship

$$\frac{10^{\frac{P_{RF(dBm)}}{10}}}{1000}(\mathrm{W}) = \frac{V_{rms}{}^2}{R} = \frac{\left(\frac{V_{RF\,peak}}{\sqrt{2}}\right)^2}{50} = \frac{V_{RF}^2}{100} \qquad (9)$$

with $P_{RF}$ in dBm being the real RF power measured with the RF power meter during the calibration step of the RF source. Despite the high resolution of the OSA, we restricted the analysis to modulation frequencies above 100MHz. For lower frequencies and low RF power, the sidebands are visible but still in the pedestal of the carrier main peak and thus their powers might be wrong as illustrated in Fig.4.

In the Fig.5, we reported the values of the Vπ voltage obtained from the fits of the curves of Fig.2 and Fig.3. Both measurement methods agree well and also agree with the manufacturer specifications, validating our methods, especially our calibration procedure.

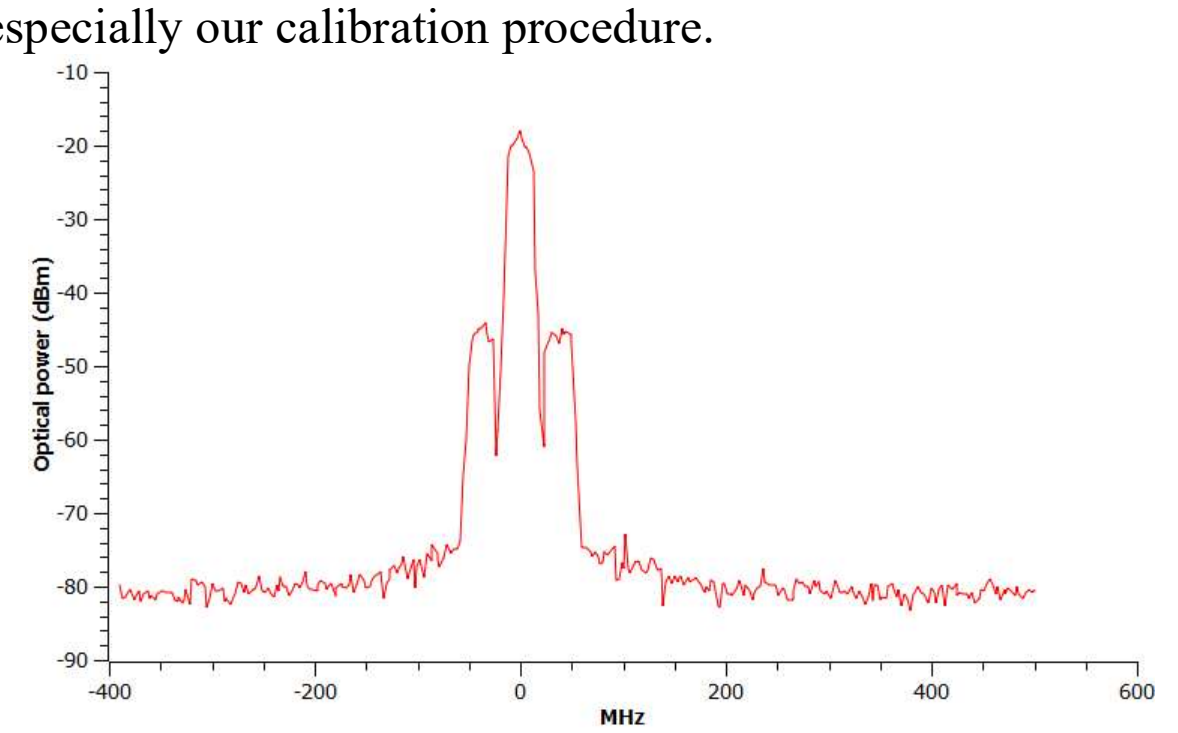

Fig.4 OSA spectrum for RF frequency of 50MHz and RF power of -25 dBm

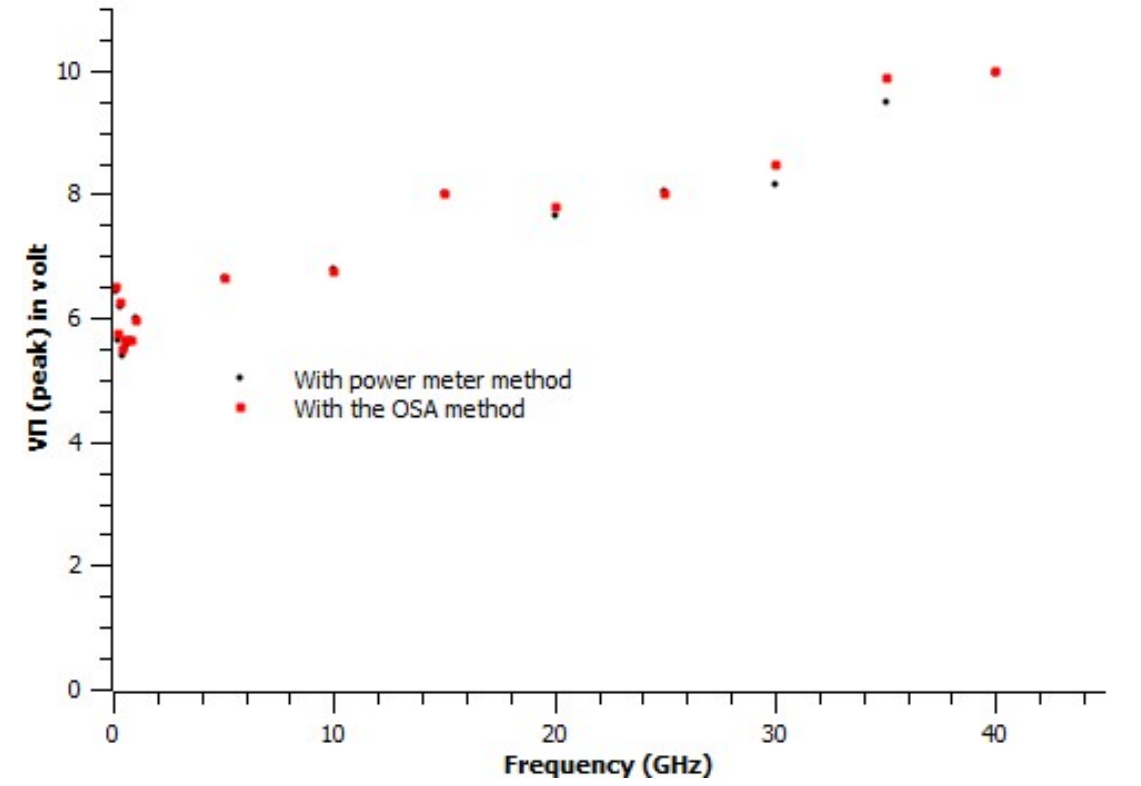

Fig.5: Vπ (peak) voltage measured with the power meter and OSA methods

## VI. Conclusions

We presented and cross-checked simple methods to measure modulator Vπ based on a power meter and an OSA. Our OSA approach is close to that of [2] in which the authors also concluded that cancelling the optical carrier (our condition (8)) to measure Vπ was too RF power demanding. They proposed a 3-time-lower-power-hungry approach by balancing the optical powers of the carrier and the first sidebands thru an optimization for every RF frequency, a process not easily automatable on the contrary to our methods that were done automatically as they require merely reading. They also considered the ratio of the optical powers of the carrier by that of the first sidebands to avoid our calibration but because variations of the optical carrier power are in the second order vs RF power (Section III), this might bring some uncertainties. We think our methods are more sensitive since they rely on measurements from noise floor: either the low level of light transmitted at the extinction of the modulators (power meter method) or the OSA noise floor. Actually, they are still reliable at 40GHz (no need to adapt the parameters $P_0$ , $\varepsilon$ or $P_{OSA,carrier}$ and both methods still agree) even though the maximum available RF peak voltage, ≈0.8V, is more than 10 times lower than Vπ, a case for which methods of [2] would not be suitable. More recently in [11], two (potentially expensive) RF generators for two-tone modulation, modulation of the DC bias, a slow photodiode and an ESA were used to avoid a fast photodiode tedious calibration. Compared to [11], we believe our approaches are much simpler with not so complicated calibration procedures that showed to be stable during the measurement duration (limited by the speed of the APEX OSA). Besides modulators with poor extinction ratio, our methods can be applied to other modulator geometries (for example modulators with two 2x2 directional couplers, one arm only being modulated…by simply adapting (2) and (5) and following the same derivations. For example, it can be readily verified that the same expression (5) can be obtained for the bar output of a MZM incorporating two 2×2 directional couplers with an intensity coupling ratio of κ.

For phase modulators, our method with an OSA is still adapted. Also, high resolution modern coherent OSAs allow retrieving more information than a power meter (intermodulation distortion [12] for optical links). If eye-diagram are not necessary, because OSA can cover several ten's of nanometer and powermeter are not limited by RF frequency, our methods are very well adapted for very high frequency considering the cost of a very fast photodiode, a suitable high bandwidth RF cable and an ESA.

## Acknowledgment

This work was supported by the European Union's Horizon 2020 under grant agreement No 101016138 (ELENA project), the Horizon Europe programme under grant agreement No 101070506, the Swiss State Secretary for Education, Research and Innovation and Innovate UK (PATTERN project). We would like to thank Nadège Courjal from Femto-ST, Besancon, France, for fruitful discussions.

## References

[1] Le Kernec, A., Sotom, M., Bénazet, Maignan, M., ... & Karafolas, N. (2017, November). Space evaluation of optical modulators for microwave photonic on-board applications. In International Conference on Space Optics—ICSO 2010 (Vol. 10565). SPIE.

[2] Shi, Y., Yan, L., & Willner, A. E. (2003). High-speed electrooptic modulator characterization using optical spectrum analysis. Journal of lightwave technology, 21(10), 2358.

[3] Soldano, L. B., & Pennings, E. C. (2002). Optical multi-mode interference devices based on self-imaging: principles and applications. Journal of lightwave technology, 13(4), 615-627.

[4] Jacobi–Anger expansion - Wikipedia

[5] Bessel Function of the First Kind -- from Wolfram MathWorld or Series Expansion of Bessel Function of the First Kind - ProofWiki

[6] De-embedding — scikit-rf Documentation

[7] De-Embedding and Embedding S-Parameter Networks Using a Vector Network Analyzer | Keysight

[8] De-embedding Techniques in Advanced Design System | Keysight

[9] Zeng, W., Wang, Y., & Wang, Z. (2025). A de-embedding method based on combining time and frequency domains. Scientific Reports, 15(1), 18637.M. Young, The Technical Writer's Handbook. Mill Valley, CA: University Science, 1989.

[10] Hawkins, R. T., Jones, M. D., Pepper, S. H., & Goll, J. H. (2002). Comparison of fast photodetector response measurements by optical heterodyne and pulse response techniques. Journal of lightwave technology, 9(10), 1289-1294.

[11] Zhang, S., Zhang, C., Wang, H., Zou, X., Liu, Y., & Bowers, J. E. (2016). Calibration-free measurement of high-speed Mach–Zehnder modulator based on low-frequency detection. Optics Letters, 41(3), 460-463.

[12] Cox III, C. H. (2006). Analog optical links: theory and practice. Cambridge University Press